\documentstyle[prbbib,twocolumn,aps]{revtex}
\begin{document}
\tolerance 50000

\draft
\twocolumn[\hsize\textwidth\columnwidth\hsize\csname @twocolumnfalse\endcsname
\title{Two interacting Hofstadter butterflies}
\author {Armelle Barelli, Jean Bellissard, Philippe Jacquod$^{a}$ and Dima L.
Shepelyansky$^{b}$}
\address {Laboratoire de Physique Quantique, UMR 5626 du CNRS, Universit\'e
Paul
Sabatier, F-31062 Toulouse Cedex, France\\
$^{a}$ Institut de Physique, Universit\'e de Neuch\^atel, CH-2000 Neuch\^atel,
Conf\'ed\'eration Helv\'etique}
\date{\today}
\maketitle

\begin{abstract}
\begin{center}
\parbox{14cm}{The problem of two interacting particles in a quasiperiodic
potential is addressed. Using analytical and
numerical methods, we explore the spectral properties and eigenstates
structure from the weak to the strong interaction case. More precisely,
a semiclassical approach based on non commutative geometry techniques permits
to understand the intricate structure of such a spectrum. An interaction
induced localization effect is furthermore emphasized. We discuss the
application of our results on a two-dimensional model of two particles in a
uniform magnetic field with on-site interaction.}
\end{center}
\end{abstract}

\pacs{
\hspace{1.9cm}
}
\vskip1pc]
\section{Introduction}
The study of crystal electrons submitted to a magnetic field
has been extensive since the early works of Landau \cite{Lan} and
Peierls \cite{Pei}. These studies have led to deep insights in the
physics of electrons in solids (interpretation of the de Haas van Alphen effect
\cite{HaAl}, investigation of the Fermi surface...). The number of
contributions
on the subject between 1950 and 1970 reveals the importance of magnetic field
effects. Twenty years ago, Hofstadter numerically computed the spectrum of the
Harper
model \cite{Har} and discovered its fractal structure as a function of the
normalized magnetic flux per lattice cell \cite{Hof} (Fig. 1).

\vspace{3mm}

\vbox to 180bp {\vfil
\centerline{\hbox to 280bp
{\includegraphics{fig1.ps}\hfil}}}

\vspace{3mm}

FIG. 1. Hofstadter's butterfly for rational values of $ \alpha =p/q $ up to $
q=29 $. For each value of the magnetic flux
$ \alpha =p/q $, we generally have $ q $ bands. Near energies equal to $ \pm 4
$ and zero flux, we observe the emergence of Landau levels.

\vspace{3mm}

The problem of a two dimensional electron on a periodic lattice has been of
special interest in solid state physics during the last fifteen years :
superconducting \cite{PaChRaVi} and normal-metal networks \cite{PaChRa:86}.
Harper-like models have been used to describe the
quantum Hall effect \cite{KoKaTa} in organic conductors, in Anyon
superconductivity \cite{ChWiWiHa} and in flux phases for the
Hubbard model \cite{RaBe:Gro}.

If the lattice is given by the positions of the ions of a metal, the lattice
spacing $ a $ is of the order of 1 \AA , so that even
with the highest magnetic fields that can be produced now, namely $ B\approx
20 T $, we get $ \alpha =\gamma /2\pi\approx 0.5\times 10^{-4} $ which is
fairly
small and shows that in this situation a ``semiclassical'' approximation will
always be relevant. As a matter of fact, an effective
Planck's constant denoted
by $ \gamma $ proportional to the applied magnetic field naturally appears as
an
adjustable variable of the
problem. Therefore the weak magnetic field limit $ \gamma\mapsto 0 $
corresponds to the semiclassical limit $ \hbar\mapsto 0 $. The corresponding
classical phase space at $ B=0 $ is the quasimomentum space, namely the
Brillouin zone of the corresponding lattice. Topologically it is a 2-torus and
the appearance of the magnetic field transforms it into a non commutative
2-torus \cite{Bel:Eva}.

Whenever $ \gamma =2\pi p/q $, ($ p,q\in {\bf N} $) the lattice Hamiltonian $ H
$ recovers some periodicity and Bloch's theory applies. We shall see
then that $ H $ can be represented as a self-adjoint $ q\times q $ matrix whose
entries are periodic functions of the quasimomentum components. Thus, if $
\gamma $ is close to any rational multiple of $ 2\pi $, it is possible to
compute the spectrum using semiclassical methods.

Based on these remarks, many theoretical and mathematical works were published
during the last fifteen years using renormalization group analysis
\cite{Wil:Cri} and pseudodifferential operators techniques \cite{HeSj:Ha1}. On
the basis of the techniques of non commutative geometry
\cite{Bel:Eva}, another approach was developed in order to reformulate and
extend the semiclassical results \cite{RaBe:Alg}.
The algebraic semiclassical approach is justified by the simplicity of its
application and its efficiency, for example in the computation of Landau
levels both in Harper-like models \cite{BaFl} and in a model-Hamiltonian
on a triangular lattice \cite{BeKrSe}. The comparison between semiclassical
formulae and exact calculations extracted from the various spectra for $
\gamma\in 2\pi {\bf Q}
$ gives surprisingly accurate agreement even for relatively large $ \gamma $'s
(namely $ \gamma /2\pi \leq 0.2 $).

While in the above formulation of the problem of Bloch electrons in a magnetic
field the particles are considered on a two-dimensional lattice, it is possible
to map it exactly onto a one-dimensional lattice with quasiperiodic potential.
The interesting property of such a lattice is the duality between momentum and
spatial coordinates pointed out by Aubry and Andr\'e \cite{AuAn}. This Aubry
duality results in a delocalized structure of the eigenstates characterized by
an algebraic decay and a multifractal eigenspectrum. This leads to a
quasidiffusive wave packet spreading on such a lattice \cite{GeKePe,WiAu}.

Recently, numbers of authors have followed a new path in the study of the
combined effect of interaction and disorder. The a priori simple problem of
two interacting particles in a random potential \cite{TIP} has indeed revealed
an unsuspectly large interaction induced delocalization effect. However, the
opposite effect has been discovered in the case of two particles in a
quasiperiodic potential. In this case, the interaction leads to the emergence
of a pure-point component out of the spectrum of the non interacting problem.
These facts have been firmly established by
overconvincing numerical and analytical results \cite{BaBeJaSh,Shepe96}. It is
one of the purposes of this paper to again express these arguments in more
details.

We shall present in this work analytical and
numerical results derived from the two-particle Harper problem with on-site
interaction on a one-dimensional lattice. More precisely we devote the second section to the presentation
of the algebraic semiclassical approach on the non interacting problem $ U=0
$. The corresponding spectrum is somehow an intricate superposition of two
Hofstadter butterflies. The aim of section 3 is to study the small interaction
regime where usual perturbation theory can be applied. The evolution of the
spectrum as a function of the strength of the interaction will be presented.
After building the analytical framework in section 4, we apply it to the
computation of the levels in the strong interaction regime. We show that for
very large $ U $, the spectrum is divided into two parts : one corresponding
to the non interacting case and the second one, looking like a Mathieu
spectrum corresponding to localized states strongly influenced by the
interaction. Based upon Aubry's duality \cite{AuAn}, it can be proved that all
the wave functions are localized in this regime as far as the Mathieu part of
the spectrum is concerned. Finally, we discuss in
section 5 the problem of two interacting particles on a two-dimensional lattice
submitted to a magnetic flux.
\section{Non interacting model}
In his 1930's study of the electronic diamagnetism of metals,
Landau computed the energy spectrum of a free electron
subject to a uniform magnetic field \cite{Lan}. If $ B $ is uniform and
parallel to one axis, for
example axis 3, the kinetic energy is written as :
\begin{eqnarray}
 H_{L}=\frac{\hbar ^{2}}{2m_{\rm e}}\left( \tilde{K}_{1}^{2}+\tilde{K}_{2}^{2}
\right)
\end{eqnarray}
with $ \tilde{K}_{\mu }=\left( P_{\mu }-q_{\rm e}A_{\mu }\right) /\hbar $, $
\mu =1,2 $ and $ A=(A_1,A_2) $ is the vector potential satisfying $ {\rm
curl}(A)=B $, $ q_{\rm e} $ is the electron charge. Moreover, the quasimomenta
$ \tilde{K}_{1},\tilde{K}_{2} $ satisfy
\begin{eqnarray}
 [\tilde{K}_{1},\tilde{K}_{2}]=iq_{\rm e}B/\hbar
\end{eqnarray}
Let us notice that this commutation rule becomes canonical when replacing $
\hbar $ by $ q_{\rm e}B/\hbar $. This new effective Planck constant (divided by
$ 2\pi
$) is proportional to the magnetic field $ B $ and behaves as a varying
physical parameter, quite naturally.

The spectrum of $ H_{L} $ is $ E_{n}=E_{0}\hbar _{{\rm eff}}\omega (2\nu +1) $
with
$ E_{0}=\hbar ^{2}/2m_{\rm e} $, $ \hbar _{{\rm eff}}=q_{\rm e}B/\hbar $ and
$ \omega =1 $. Therefore :
\begin{eqnarray}
 E_{\nu }=\hbar\omega _{c}(\nu +1/2)
\end{eqnarray}
where $ \omega _{c}=q_{\rm e}B/m_{\rm e} $ is the cyclotronic frequency and $
\nu $ is the Landau quantum number.

When $ B=0 $, the electron energy $ E(k) $ for each
conduction band is given by Bloch's theory, where the quasimomentum components
$ k=(k_{1},k_{2}) $ are defined modulo
the reciprocal lattice such that for a simple square lattice in the
tight-binding approximation $ E(k)=2E_{0}\left[\cos (k_{1}a_{1})+\cos
(k_{2}a_{2})\right] $ where $ a_{\mu } $ is the vector of the Bravais lattice
in the $ \mu
$-direction. The charge carriers energy is calculated by expanding $
E(k) $ near its extremum, denoted by $ k_{c} $, namely :

\begin{eqnarray}
 E(k)=E(k_{c})+ \hbar ^{2}\left( M^{-1}\right) _{ij}
k_{i}k_{j}/2+{\rm O}\left( \vert k\vert ^{3}\right)
\end{eqnarray}
where $ M $ stands for the effective mass matrix such that $ M^{-1}=D^{2}E(
k_{c})/\hbar ^{2} $.

Thus Landau theory leads to a substitution $ k_{i}\cdot a_{i}\longmapsto
\hat{K}_{i}=\frac{1}{\hbar }
\left(P-q_{\rm e}A\right)\cdot a_{i} $ when an external magnetic field is
applied. We have the following commutation
rule :
\begin{eqnarray}
[\hat{K}_{i},\hat{K}_{j}]=iq_{\rm e}Ba_{i}a_{j}/\hbar =2i\pi\phi
_{ij}/\phi _{0}=2i\pi\alpha =i\gamma
\end{eqnarray}
where $ \phi _{0}=h/q_{\rm e} $ is the flux quantum, $ \phi _{ij} $ is the
magnetic flux through the cell generated by $ \left( a_{i}, a_{j}\right)
$ and $ \alpha =\phi _{ij}/\phi _{0} $ is the normalized magnetic flux. For a
crystal with periodic spacing, the Peierls operator $ {\cal P}(k) $ is
represented by an effective Hamiltonian \cite{Pei}, namely :
\begin{eqnarray}
{\cal P}(k)=\sum _{m}h_{m}(\alpha ){\rm e}^{i
m\cdot k}\mbox{ , }\hspace{0.2cm}m\in {\bf Z}^{2}
\end{eqnarray}
where $ h_{m}(\alpha ) $ are smooth functions of $ \alpha $. Thus :
\begin{eqnarray}
H_{\rm eff}(\hat{K}_{1},\hat{K}_{2})=\sum _{m}h_{m}
{\rm e}^{im\cdot \hat{K}}
\end{eqnarray}
If several bands intersect the Fermi level, the interband coupling due to the
magnetic field is neglected and therefore :
\begin{eqnarray}
H_{{\rm eff}}=2t\left(\cos \hat{K}_{1}+\cos \hat{K}_{2}\right)
\end{eqnarray}
where $ t $ is physically interpreted as a transfer term corresponding to the
required energy for an electron to jump from one site to another (nearest
neighbour) of the lattice.

For a wave function $ \psi (n_{1},n_{2}) $ defined on the two-dimensional
lattice $ \ell ^{2}({\bf Z}^{2}) $, the magnetic field effect can be seen
through the magnetic translation operators such that :
\begin{eqnarray}
\left( {\cal U}_{1}\psi\right) (n_{1},n_{2}) & = & {\rm e}^{\frac{-iq_{\rm
e}}{\hbar}
\int_{(n_{1}-
1,n_{2})}^{(n_{1},n_{2})}\vec{A}\cdot\vec{dl}}\psi (n_{1}-1,n_{2}) \nonumber
\\
\left( {\cal U}_{2}\psi\right) (n_{1},n_{2}) & = & {\rm e}^{\frac{-
iq_{\rm e}}{\hbar}\int_{(n_{1},n_{2}-1)}^{(n_{1},n_{2})}\vec{A}\cdot
\vec{dl}}
\psi (n_{1},n_{2}-1)
\end{eqnarray}
\noindent in an appropriate gauge we get :
\begin{eqnarray}
\left( {\cal U}_{1} \psi\right) (n_{1},n_{2}) & = & \psi (n_{1}-1,n_{2})
\nonumber \\
\left( {\cal U}_{2} \psi\right) (n_{1},n_{2}) & = & {\rm e}^{-i\gamma n_{1}}
\psi (n_{1},n_{2}-1)
\end{eqnarray}
Because of the presence of the uniform
magnetic field, the magnetic translation operators no longer commute, namely in
that case
\begin{eqnarray}
{\cal U}_{1}{\cal U}_{2}={\rm e}^{i\gamma }{\cal U}_{2}{\cal U}_{1}
\end{eqnarray}
where $ \gamma $ is the normalized magnetic flux per lattice-cell defined by $
\gamma =2\pi\alpha =2\pi\phi /\phi _{0} $, $ \phi $ being the flux per unit
cell and $ \phi _{0}=h/q_{\rm e} $ the flux quantum.

\noindent If we set $ {\cal U}_{1}=\exp\left( i\hat{K}_{1}\right) $, $ {\cal
U}_{2}=\exp
\left( i\hat{K}_{2}\right) $ using the commutation rule (11), we obtain
\begin{eqnarray}
[\hat{K}_{1},\hat{K}_{2}]=\frac{iq_{\rm e}Ba_{1}a_{2}}{\hbar }=2i\pi\frac{\phi
}
{\phi _{0}}=2i\pi\alpha =i\gamma
\end{eqnarray}
which corresponds to (5) in the particular case $ i=1 $ and $ j=2 $.

Following Harper \cite{Har}, the eigenvalue equation is written
\begin{eqnarray}
& & E_{0}\left[\psi (n_{1}+a,n_{2})+\psi (n_{1}-a,n_{2})+\right. \nonumber \\
& & \left. +\lambda {\rm e}^{iq_{\rm e}Bn_{1}a/
\hbar }\psi (n_{1},n_{2}+a)+\lambda {\rm e}^{-iq_{\rm e}Bn_{1}a/\hbar }\psi
(n_{1},n_{2}-a)\right]  \nonumber \\
& & \hspace{4cm} =  2{\cal E}\psi (n_{1},n_{2})
\end{eqnarray}
$ \lambda $ represents the strength of the quasiperiodic potential.

Let us assume plane-wave behaviour in one direction, i.e. we set $ \psi
(n_1,n_2)=\int d\beta {\rm e}^{i\beta n_2}\phi (n_1) $ since the coefficients in
the previous equation only involve $ n_{1} $ :
\begin{eqnarray}
\psi (n_{1},n_{2})={\rm e}^{i\beta n_{2}}\phi (n_{1}) \nonumber
\end{eqnarray}
and the eigenequation becomes :
\begin{eqnarray}
\phi (n_{1}+1)+\phi (n_{1}-1)+2\lambda\cos\left(2\pi\alpha n_{1}+\beta \right)
\phi (n_{1}) \nonumber \\
={\cal E}\phi (n_{1})
\end{eqnarray}
where we included the additive energy due to the motion in the field direction
in the eigenvalue $ {\cal E} $ and where we changed the origin of $ n_{1} $.

It is possible to characterize the properties of eigenfunctions from (14) by
looking at a special regime, namely $ \lambda\ll 1 $. Therefore, the hopping
term is dominant and we can treat the quasiperiodic potential part of the
eigenvalue equation as a perturbation. It is then easy to see that the
solutions are given for $ \lambda =0 $ by Bloch waves $ \phi _{k}(n )=\exp
(ikn) $ with an energy $ E=2\cos k $. For $ 0<\lambda\ll 1 $, the perturbation
theory allows us to perform an expansion of eigenvalues and eigenstates in $
\lambda $ such that
\begin{eqnarray}
E(k)&=&2\cos k+\sum _{m}\lambda ^m\epsilon _m(k)\nonumber\\
\phi _{k}(n)&=&{\rm e}^{ikn}\left(1+\sum _{m}\lambda ^m f_{m}(\gamma n+\beta
)\right)\nonumber\\
&=&{\rm e}^{ikn}u_{m}(\gamma n+\beta )
\end{eqnarray}
Evaluating the first and second order perturbation theory contributions and
replacing the expressions (15) in (14) leads to :
\begin{equation}
\lambda\left(u_{m+1}+u_{m-1}\right)+2\cos (\gamma m+k)u_m=E(k)u_m
\end{equation}
The previous equation is known as the ``almost Mathieu'' eigenvalue equation
and the argument above is the Aubry duality \cite{AuAn} between momentum and
coordinate representations. As far as spectral properties are concerned one can
be easily convinced that dealing with Bloch states means that the states are
extended. Thanks to this duality, $ \lambda\leftrightarrow 1/\lambda $ between
(14) and (16), it is quite natural to get localized states for the almost
Mathieu Hamiltonian at small $ \lambda $'s. More precisely, it has been proved
that the spectrum of the almost Mathieu Hamiltonian is pure-point at small
$ \lambda $'s and for almost all $ \beta $'s \cite{BeLiTe}. Conversely if
$ \lambda\gg 1 $, the almost Mathieu Hamiltonian has purely continuous spectrum
for almost all $ \beta $'s \cite{Cho}.

Setting $ t=1 $ in formula (8) and using the magnetic translation
operators $ {\cal U}_{1} $ and $ {\cal U}_{2} $ defined on the two-dimensional
square
lattice by (10), the previous Harper equation can be written as the action
of an effective Hamiltonian such that :
\begin{eqnarray}
H_{{\rm eff}}={\cal U}_{1}+{\cal U}_{1}^{-1}+{\cal U}_{2}+{\cal U}_{2}^{-1}
\end{eqnarray}

In order to study the two interacting particles model on a quasiperiodic
lattice we
transform the previous eigenvalue equation (14) into ($\lambda =1$)
\begin{eqnarray}
& & \left[ 2\cos\left(\gamma n_{1}+\beta _1 \right)+2\cos\left(\gamma
n_{2}+\beta _2 \right)+U\delta _{n_1,n_2}\right]\phi _{n_{1},n_2} \nonumber \\
& & +\phi _{n_{1}+1,n_2}+\phi _{n_{1}-1,n_2}+\phi _{n_{1},n_2+1}+\phi
_{n_{1},n_2-1} \nonumber \\
& & \hspace{5cm} =E\phi _{n_{1},n_2}
\end{eqnarray}
where $ \beta _{1,2} $ are related to the quasimomentum components of the non
interacting case. In the following we shall consider $ \beta _{1,2}=\beta $. 
Here we chose the form of on-site interaction which only
influences the symmetric configurations while the antisymmetric ones remain not
affected by $ U $. Due to that, we shall only discuss symmetric
configurations in the following.

In the most simple case of non interacting particles ($ U=0 $), the spectrum
can be computed as before and is shown in Fig. 2.

\vspace{3mm}

\vbox to 180bp {\vfil
\centerline{\hbox to 280bp
{\includegraphics{fig2.ps}\hfil}}}

\vspace{2mm}

FIG. 2. Spectrum of the two-particle Harper problem with $ U=0 $ obtained for
rational values of $ \alpha =p/q $ up to $ q=19 $.

\vspace{3mm}

As we pointed out before, $ \gamma =
2\pi\alpha $ appears in our problem as an effective Planck constant since the
magnetic
translation operators $ {\cal U}_{1} $ and $ {\cal U}_{2} $ obey canonical
commutation rules in $ \gamma $. Therefore, we study the semiclassical
limit by letting $ \gamma\mapsto 0 $. It is also possible to perform
calculations near a rational value of the magnetic flux, namely $ \gamma
'=\gamma -2\pi p/q\longmapsto
0 $. The efficiency and the accuracy of our calculations allow to explain some
features of the corresponding spectra.

\noindent When $ \gamma =0 $, corresponding to $ B=0 $, we recover the
band function $ E(k) $, where $ k=(k_{1},k_{2}) $. To study the Landau levels,
we expand the classical symbol of
the Hamiltonian around an extremum of the band function denoted by $ k_{c} $ :
\begin{eqnarray}
{\cal H}(k)={\cal H}(k_{c})+\frac{1}{2}\partial _{\mu }
\partial _{\nu }{\cal H}(k_{c})k_{\mu }k_{\nu }+\cdots
\end{eqnarray}
The quantization of $ {\cal H}(k) $ consists in replacing the
magnetic translation operators by \cite{RaBe:Alg}:
\begin{eqnarray}
{\cal U}_{j}=\exp\left( i(k_{cj}+\sqrt{\gamma }K_{j})\right) \mbox{ , } j=1,2
\end{eqnarray}
where $ k_{cj} $ are the bottom well coordinates and $ K_{j} $ are operators
satisfying Heisenberg's commutation relations $ [K_1,K_2]=i $. The quantized of
$ {\cal H } $, denoted by $ H $, is written as :
\begin{eqnarray}
H=\sum _{m}h(m,\gamma ){\rm e}^{i(m\cdot k
_{c}+\sqrt{\gamma }m\cdot K)} \nonumber
\end{eqnarray}
with $ m\cdot K=m_{1}K_{1}+m_{2}K_{2} $. In the weak field limit, one formally
expands $ H $ in powers of $ \sqrt{\gamma } $ :
\begin{eqnarray}
H&=&\sum _{m}\left\{ h(m,0){\rm e}^{im\cdot
k_{c}}+i\sqrt{\gamma }h(m,0)
{\rm e}^{im
\cdot k_{c}}m\cdot K\right. \nonumber \\
&&\left. +\gamma \left[
\frac{\partial h}{\partial
\gamma }(m,0){\rm e}^{im\cdot k_{c}}-\frac{1}{2}
h(m,0){\rm e}^{im\cdot k_{c}}(m\cdot K)
^{2}\right]\right\} \nonumber \\
&&+\mbox{O}(\gamma ^{3/2})
\end{eqnarray}
which we rewrite as :
\begin{eqnarray}
H&=&{\cal H}(k_{c},0)+\gamma \left(\frac{\partial {\cal H}}{\partial\gamma
}(k_{c},0)-
\frac{1}{2}\partial _{\mu }\partial _{\nu }{\cal H}(k_{c},0)K_{\mu }
K_{\nu }\right) \nonumber \\
&&+\mbox{O}(\gamma ^{3/2})
\end{eqnarray}
\noindent The $ \partial {\cal H}/\partial\gamma $-term takes into
account a possible explicit $ \gamma $-dependence of the classical Hamiltonian
whereas
$ \partial _{\mu }\partial _{\nu }{\cal H} $ represents the inverse effective
mass matrix due
to the band function curvature. By a unitary transformation, the quadratic term
can be written as
$ \omega \left( {\cal K}_{1}^{2}+{\cal K}_{2}^{2}\right) /2 $ where
$ \omega $ is related to the determinant of the Hessian matrix $ \partial _{\mu
}\partial
_{\nu } {\cal H}(k_{c},0) $. We recognize here
the harmonic oscillator Hamiltonian. For this reason, the energy levels denoted
by $
E_{\nu } $ are called ``Landau levels'' and are equal, to that order in $
\gamma $, to $ \omega (\nu +1/2) $ leading to :
\begin{eqnarray}
E_{\nu }(\gamma )&=&{\cal H}(k_{c},0)+\gamma (2\nu +1)\left( \det\frac{1}{2}
D^{2}{\cal H}(k_{c},0)\right)^{1/2} \nonumber \\
&&+\gamma \left(\frac{\partial {\cal H}
(k_{c},0)}{\partial\gamma }\right)+\ldots +{\rm O}(\gamma ^{N})
\end{eqnarray}

The formula (23) has been checked numerically on several
models. To illustrate it, let us consider the two-particle
Harper Hamiltonian on the square lattice (18) near the maximum $ k_{c}=(0,0) $
of the band function. Using
(20) and (22) the quantized Hamiltonian is then expressed as an expansion
in powers of $ \gamma $ :
\begin{eqnarray}
H&=&8-\gamma\left((K_{1}^{(1)})^{2}+(K_{2}^{(1)})^{2}+
(K_{1}^{(2)})^{2}+(K_{2}^{(2)})^{2}\right) \nonumber \\
&&+\frac{\gamma^{2}}{3}\left((K_{1}^{(1)})^{4}+(K_{2}^{(1)})^{4}+(K_{1}^{(2)})^{4}+(K_{2}^{(2)})^{4}\right) \nonumber \\
&&+{\rm O}(\gamma ^{3})
\end{eqnarray}
where the $ K^{(1,2)} $ are quasimomenta for particle 1 and 2 respectively.
Finally it gives the Landau levels :
\begin{eqnarray}
E_{\nu _{1},\nu _{2}}(\gamma )&=&8-2\gamma (\nu _1+\nu _2+1) \nonumber \\
&&+\gamma ^{2}\left[(2\nu _1+1)^{2}+(2\nu _2+1)^{2}+2\right]/16 \nonumber\\
&&+{\rm O}(\gamma ^{3})
\end{eqnarray}
where $ \nu _1 $ and $ \nu _2 $ are the Landau quantum numbers associated with
particle 1 and 2 respectively. To check the accuracy of this formula, we
compared it to the datas extracted from the numerical spectrum obtained by
exact diagonalization. Fig. 3 shows the accuracy of such a semiclassical
expansion in the description of the spectrum of the two-particle Harper model
when $ \gamma\mapsto 0 $.

\vspace{-6mm}
\let\picnaturalsize=N
\def\picsize{4.0in}
\def\picfilename{fig3.eps}
\ifx\nopictures Y\else{\ifx\epsfloaded Y\else\input epsf \fi
\let\epsfloaded=Y
\centerline{\ifx\picnaturalsize N\epsfxsize \picsize\fi
\epsfbox{\picfilename}}}\fi

\vspace{-1cm}

FIG. 3. Comparison between semiclassical calculations (25) (full curves) and
exact numerical spectrum (points) for Landau sublevels in the two-particle
Harper model on the square lattice when $ U=0 $. Datas are extracted in the
region of energies corresponding to the maximum $ (0,0) $ of the band function.

\vspace{3mm}

\section{Weak interaction regime}

We present here a simple perturbative treatment that enables to implement the
already presented
results for the weakly interacting case. The first-order contribution allows to
understand the splitting of Landau bands at sufficiently weak interaction, and
describes it qualitatively well. It moreover enlightens the mechanism through
which interaction affects the system.
Using the representation defined by (20), we write the unperturbed Hamiltonian
as :
\begin{equation}
H_{\rm eff} = 2 \cos(\sqrt{\gamma} K_1) + 2 \cos(\sqrt{\gamma}K_2)
\end{equation}

In the semiclassical limit $\gamma \mapsto 0$ we expand (26) in a power series
around a minimum of potential $q_N=\pi/\sqrt{\gamma}+2\pi N/\sqrt{\gamma}$, $N
\in {\bf Z}$. Keeping only terms up to the second order in $\gamma$ we
end up with a harmonic oscillator. In this approximation and in the continuous
case the one-particle wave functions of the unperturbed Hamiltonian are
therefore given by :
\begin{eqnarray}
\psi_{\nu}(y)  = H_{\nu}(\frac{y}{\sqrt{\gamma}})  \exp
\left(-\frac{y^2}{2\gamma}\right)/{\sqrt{2^{\nu} \nu! \sqrt{ \gamma \pi}}}
\end{eqnarray}
Here, $H_{\nu}(x)$ is a Hermite polynomial, the index $\nu$ refers to the
Landau level, $y= x-q_{N}$ in term of the minimum of potential $q_{N}$ around
which the harmonic approximation has been performed, and $x$ is the spatial
coordinate. This expression is of course valid, provided $ \gamma $ and
$|x-q_{N}| \ll 1$, i.e. in the small magnetic field regime, and not too far
away from a potential minimum. Extending our expansion to higher powers in
$\gamma$ would allow us to increase the range of validity of this expression.
We could indeed write the exact normalized wave functions in an expansion in
$\gamma$ as
\begin{equation}
\varphi_{\nu}(y) = \exp \left( -\frac{y^2}{2\gamma}\right) ( c_0
H_{\nu}(\frac{y}{\sqrt{\gamma}}) + \gamma c_1
H_{\nu}^{(1)}(\frac{y}{\sqrt{\gamma}}) + ...)
\end{equation}

For the purpose of discretization, we introduce a continuous variable $\xi \in
{\bf R}$ labeling the well, and a discrete one $l \in {\bf Z}$ numbering the
sites. Then $y=\xi-l \sqrt{\gamma}$ since in the chosen representation, the
intersite spacing is $a=\sqrt{\gamma}$. The set $\{\varphi_{\nu}\}$ builds a
quasiorthogonal basis in the sense that for $\xi \neq \xi'$, due to the
Gaussian envelop of the states we have :
\begin{eqnarray}
\sum_l  \varphi_{\nu}(\xi-l\sqrt{\gamma}) \varphi_{\mu}(\xi'-l\sqrt{\gamma}) =
{\rm O} (\exp(-1/\gamma)) \delta_{\mu,\nu}
\end{eqnarray}
These functions are periodic in $\xi$ with period $1/\sqrt{\gamma}$. In the
semiclassical limit the norm of $\varphi_{\nu}$ is :
\begin{eqnarray}
\| \varphi_{\nu} \|^2 & = & \sum_{l=-\infty}^{\infty}
|\varphi_{\nu}(\xi-l\sqrt{\gamma})|^2  \nonumber \\
& = &  1/\sqrt{\gamma} \int dy |\varphi_{\nu}(y)|^2 \nonumber \\
& = & 1/\sqrt{\gamma}
\end{eqnarray}
Consequently, to get normalized one-particle wave functions on the discrete
lattice $\ell ({\bf Z})$ we must multiply the $\varphi$'s by a factor
$\gamma^{1/4}$. We thus can write the symmetrized two-particle unperturbed wave
functions as :
\begin{eqnarray}
\phi_{\nu,\mu}^{\xi,\xi'} (l,l') = \sqrt{\frac{\gamma}{2}} \left(
\varphi_{\nu}(\xi-l \sqrt{\gamma}) \varphi_{\mu}(\xi'-l' {\gamma}) \right.
\nonumber \\
\left. \pm \varphi_{\mu}(\xi-l \sqrt{\gamma}) \varphi_{\nu}(\xi'-l' {\gamma})
\right) \left(1-\delta_{\mu,\nu} (1-1/\sqrt{2}) \right)
\end{eqnarray}

We are now able to compute the first-order correction for the energy. Because
of the exponentially localized character of (28), two particles located on
different wells have only an exponentially small overlap, and as a consequence
do practically not interact. Therefore the first-order interaction induced
correction to the energy is non zero only for symmetric wave functions with
$\xi = \xi'$. We have :
\begin{eqnarray}
\Delta E^{(1)} & = & U \sum_l \sum_{l'} (\phi_{\nu,\mu}^{\xi,\xi'} (l,l'))^2
\delta(\xi-\xi'+(l'-l) \sqrt{\gamma}) \nonumber \\
& = & U \delta_{\xi,\xi'} \int dy \left(\varphi_{\mu} (y) \varphi_{\nu} (y)
\right)^2 (2-\delta_{\mu,\nu}) + \nonumber \\
& & + {\rm O}(\exp(-1/\gamma))
\end{eqnarray}

\vspace{-8mm}

\let\picnaturalsize=N
\def\picsize{4.0in}
\def\picfilename{fig4.eps}
\ifx\nopictures Y\else{\ifx\epsfloaded Y\else\input epsf \fi
\let\epsfloaded=Y
\centerline{\ifx\picnaturalsize N\epsfxsize \picsize\fi
\epsfbox{\picfilename}}}\fi

FIG. 4. Spectrum of the two-particle Harper model with on-site interaction at
$U=0.4$ up to $ q=23 $.

\vspace{3mm}

{}From (28), the dominant term in the last integral is of order
O$(\sqrt{\gamma})$ so that we finally have
\begin{eqnarray}
\Delta E^{(1)} & \sim &  U \delta_{\xi,\xi'} \sqrt{\gamma}
\end{eqnarray}
The numerical factor can be estimated from the harmonic approximation (27)
which leads to :
\begin{eqnarray}
\Delta E_{h}^{(1)} & = &  U \delta_{\xi,\xi'} \sqrt{\frac{\gamma}{2 \pi}} =  U
\delta_{\xi,\xi'} \sqrt{\alpha}
\end{eqnarray}
for states with Landau quantum numbers (0,0) and (0,1).
This result shows that the interaction primarily acts on two-particle states
with high double-site occupancy. In what follows we shall call such states
``pair states''. States for which the particles are located around different
potential minima practically do not feel the interaction. Therefore, switching
on the interaction does not modify most of the spectrum as can be seen on Fig.
4.

{}From (25) and (34) and for small enough interaction, the shifted part of the
spectrum is given by :
\begin{eqnarray}
E_{\nu _{1},\nu _{2}}(\gamma )& \sim &8+ U  \sqrt{\frac{\gamma}{2 \pi}}-2\gamma
(\nu _1+\nu _2+1) \nonumber \\
&&+\gamma ^{2}\left[(2\nu _1+1)^{2}+(2\nu _2+1)^{2}+2\right]/16
\end{eqnarray}

The amazing agreement between the numerically computed spectrum obtained by
exact Lanczos diagonalization and (35) is shown in Fig. 5 where $U=0.4$. It is
a confirmation of our reasoning : pair states form the shifted part of the
spectrum. Because these states are much fewer than states where particles are
located in different wells, the shifted spectrum is much less dense. In this
sense the interaction splits the butterfly into two parts. One of them is 
practically not affected by the interaction and corresponds to the states 
where particles are far from each other. The second one is shifted and relays 
to the situation where particles form pair states. Here, the interaction 
results in a global shift of the spectrum. In this way, new
states appear in the initial gaps of the non interacting spectrum (see Figs.
4,6 and Fig. 1(b) in \cite{BaBeJaSh}). Direct
analysis of eigenstates shows that the corresponding states are exponentially
localized \cite{BaBeJaSh}. We shall come back to this point later on for the
case of strong interaction.

\let\picnaturalsize=N
\def\picsize{4.0in}
\def\picfilename{fig5.eps}
\ifx\nopictures Y\else{\ifx\epsfloaded Y\else\input epsf \fi
\let\epsfloaded=Y
\centerline{\ifx\picnaturalsize N\epsfxsize \picsize\fi
\epsfbox{\picfilename}}}\fi

\vspace{-1cm}
FIG. 5. Comparison between semiclassical calculations extended by perturbation
theory (35) (full curves) and exact numerical spectrum (points) for the
two-particle Harper model with on-site interaction at $U=0.4$.

\vspace{3mm}

\section{Strong interaction regime}
The strongly interacting regime needs a special treatment quite analogous to
the one presented in section 2. As we will see, Schur's complement formula can
be successfully applied to construct an effective Hamiltonian. The latter is
then expanded in a power series in $\gamma$ to deliver highly accurate
formulae. From the weakly interacting regime we learned that particles located
on different potential minima do not feel each other : for such pairs, the
interaction is suppressed by an exponentially small term of order ${\rm O}
(U \exp(-1/\gamma))$. Therefore, this picture remains valid even for large
$U$'s, the relevant parameter being the magnetic flux. Pair states on the other
hand undergo an energy increase of order $\Delta E \approx U$. Therefore when
the strength of the interaction $ U>0 $ increases, one part of the spectrum is
almost not affected. Another spectral structure appears, initially looking like
a shifted butterfly (see Fig. 4 where $ U=0.4 $), then evolving to a shifted
Mathieu spectrum as the interaction grows bigger and bigger (see Figs. 6, 7 and
8
where $U=5,10 $ and $20$ respectively).

\let\picnaturalsize=N
\def\picsize{4.0in}
\def\picfilename{fig6.eps}
\ifx\nopictures Y\else{\ifx\epsfloaded Y\else\input epsf \fi
\let\epsfloaded=Y
\centerline{\ifx\picnaturalsize N\epsfxsize \picsize\fi
\epsfbox{\picfilename}}}\fi

FIG. 6. Spectrum of the two-particle Harper model in the intermediate regime
$ U=5 $ up to $ q=23 $.

\let\picnaturalsize=N
\def\picsize{4.0in}
\def\picfilename{fig7.eps}
\ifx\nopictures Y\else{\ifx\epsfloaded Y\else\input epsf \fi
\let\epsfloaded=Y
\centerline{\ifx\picnaturalsize N\epsfxsize \picsize\fi
\epsfbox{\picfilename}}}\fi

FIG. 7. Spectrum of the two-particle Harper model in the strongly interacting
regime $ U=10 $ up to $ q=23 $.

\vspace{-5mm}

\let\picnaturalsize=N
\def\picsize{4.0in}
\def\picfilename{fig8.eps}
\ifx\nopictures Y\else{\ifx\epsfloaded Y\else\input epsf \fi
\let\epsfloaded=Y
\centerline{\ifx\picnaturalsize N\epsfxsize \picsize\fi
\epsfbox{\picfilename}}}\fi

\vspace{-7mm}

FIG. 8. Spectrum of the two-particle Harper model in the strongly interacting
regime $ U=20 $ up to $ q=23 $.

\vspace{3mm}

In this section, we present an analytical approach that allows to understand
completely the mechanism driving this evolution of the spectrum. Further
details like the splitting of the Landau band $\nu_1=0,\nu_2=1$ will also be
computed, even though the physics is there less transparent (see Fig. 7). We
shall concentrate our semiclassical calculation near the band
function maximum $ k_c=(0,0) $ corresponding to the energy $ z\approx U+4 $ in
the spectrum. The two-particle Hamiltonian can be expressed in the following
way :
\begin{equation}
\begin{array}{l}
H_{\rm TIP}=\sum _{m,n}\left[ 2\cos\left(\gamma
m+\beta\right)+2\cos\left(\gamma n+\beta\right)\right]\\ \\
\vert m\otimes n\rangle\langle m\otimes n\vert
+U\sum _{m}\vert m\otimes m\rangle\langle m\otimes m\vert\\ \\
+\sum _{m\neq n}\vert m\otimes n\rangle\left[\langle m\otimes n+1\vert +
\langle m\otimes n-1\vert\right.\\ \\
\left. +\langle m+1\otimes n\vert + \langle m-1\otimes n\vert\right]
\end{array}
\end{equation}
The strategy is based on the so-called Schur complement formula. Our
Hamiltonian $ H_{\rm TIP} $ is a self-adjoint operator acting on a Hilbert
space that can be decomposed as $ {\cal H}={\cal P}\oplus {\cal Q} $. Let $ P $
and $ Q $ be the orthogonal projections on each subspace of that decomposition,
namely :
$$ \begin{array}{l}
P=\sum _{m}\vert m\otimes m\rangle\langle m\otimes m\vert\\ \\
Q={\bf I}-P=\sum_{m\neq n}\vert m\otimes n\rangle\langle m\otimes
n\vert\end{array} $$
In other words, $ P $ is the eigenprojection on pair states and $ Q $ is its
orthogonal.
If $ z $ is an eigenvalue of $ H_{\rm TIP} $ and does not belong to the
spectrum of $ QH_{\rm TIP}Q $ then it is also an eigenvalue of the following
effective Hamiltonian
\begin{equation}
H_{\rm TIP}^{{\rm eff}}(z)=P H_{\rm TIP}P +PH_{\rm TIP}Q\frac{1}{z-QH_{\rm
TIP}Q}Q H_{\rm TIP}P
\end{equation}
When $ U $ is large the dominant term in the effective Hamiltonian given by the
Schur complement formula (37) corresponds to the pair states. The semiclassical
approach we introduced in section 2 remains valid so that $ H_{\rm TIP}^{{\rm
eff}}(z)=H_{\rm TIP}^{{\rm eff}}(z_0+\gamma z_1 +\gamma ^2 z_2+{\rm O}(\gamma
^3 )) $. The implicit equation to be solved is then :
\begin{equation}
H_{\rm TIP}^{{\rm eff}}(z)=z_0+\gamma z_1 +\gamma ^2 z_2+{\rm O}(\gamma ^3 )
\end{equation}
with
\begin{equation}
H_{\rm TIP}^{{\rm eff}}(z)=H_{\rm TIP}^{(0)}(z)+\gamma H_{\rm
TIP}^{(1)}(z)+\gamma ^2 H_{\rm TIP}^{(2)}(z)+{\rm O}(\gamma ^3)
\end{equation}
The expansion of the dominant term reads :
\begin{equation}{\mbox{$ \begin{array}{rcl}
P H_{\rm TIP}P&=&U+4\cos (\sqrt{\gamma }K_2)=U+4-2\gamma K_2^2\\ \\
&&\displaystyle +\frac{\gamma ^2}{6}K_2^4+{\rm O}(\gamma ^3) \end{array} $}}
\end{equation}
and if we consider $ U $ large, $ z $ is large too so that :
\begin{equation}{\mbox{$ \begin{array}{rcl}
\displaystyle\frac{1}{z-QH_{\rm
TIP}Q}&=&\displaystyle\frac{1}{z}+\displaystyle\frac{QH_{\rm
TIP}Q}{z^2}+\displaystyle\frac{QH_{\rm TIP}QQH_{\rm TIP}Q}{z^3}\\
&&\displaystyle+{\rm O}\left(z^{-4}\right)\end{array} $}}
\end{equation}
Expressing the different contributions in Schur's formula and expanding in
powers of $ \gamma $ lead to :
\begin{equation}{\mbox{$\begin{array}{rcl}
H_{\rm TIP}^{(0)}(z)&=&\displaystyle
4+U+\displaystyle\frac{8}{z}+\displaystyle\frac{32}{z^2}+\displaystyle\frac{176}{z^3}+{\rm O}\left( z^{-4}\right)\\ \\
H_{\rm
TIP}^{(1)}(z)&=&\displaystyle\frac{-2(z^3+8z+64)}{z^3}\left[K_{2}^{2}+\frac{z^2+4z+34}{z^3+8z+64}K_1^{2}\right]\\ \\
H_{\rm TIP}^{(2)}(z)&=&\displaystyle\frac{z^3+8z+256}{z^3}\left[
K_2^4+\frac{z^2+4z+70}{z^3+8z+256}K_1^4\right]\\ \\
&&+2\displaystyle\frac{z+8}{z^3}\left(
K_1^2K_2^2+K_2^2K_1^2\right)-8\frac{z+8}{z^3}\\ \\
&&+16\displaystyle\frac{(z+8)^2}{z^3(z^3+8z+64)}\left(K_1^2+K_2^2\right)
\end{array} $}}
\end{equation}
Finally, we have to solve (38) to get the coefficients $ z_0 $, $ z_1 $ and
$ z_2 $. The corresponding equations for those coefficients are at most of
degree four. We shall give here the equation that $ z_0 $ has to satisfy at the
order $ {\rm O}\left(\displaystyle z^{-4}\right) $
\begin{equation}
4+U+\frac{8}{z_0}+\frac{32}{z_0^2}+\frac{176}{z_0^3}=z_0
\end{equation}
In a very similar way used for the computation of $ z_0 $, the analytical
expressions of $ z_1 $ and $ z_2 $ can be derived from (38), (42) and (43).
The good agreement with the exact numerical spectrum can be seen on Fig. 9 for
$ U=50 $. Here the numerical values for the sublevels are : for $ \nu _{1,2}=0
$, $ z(\gamma )=54.1597-0.2826\gamma +0.0356\gamma ^2 $, for $ \nu _{1,2}=(0,1)
$, $ z(\gamma )=54.1597-0.8480\gamma +0.2084\gamma ^2 $, for $ \nu _{1,2}=(1,1)
$, $ z(\gamma )=54.1597-1.4133\gamma +0.5539\gamma ^2 $.
A similar computation can be done near the band function minimum  $ k_c=(\pi ,
\pi ) $ corresponding to the energy $ z\approx U-4 $.

\vspace{-5mm}

\let\picnaturalsize=N
\def\picsize{4.0in}
\def\picfilename{fig9.eps}
\ifx\nopictures Y\else{\ifx\epsfloaded Y\else\input epsf \fi
\let\epsfloaded=Y
\centerline{\ifx\picnaturalsize N\epsfxsize \picsize\fi
\epsfbox{\picfilename}}}\fi

\vspace{-1cm}
FIG. 9. Comparison between semiclassical calculations (full curves, see text)
and exact
numerical spectrum (points) for levels in the two-particle Harper model for
$ U=50 $.

\vspace{3mm}

The structure of the pair states for $ U\gg 1 $ can be understood in the
following way : the diagonal term corresponding to the energy of particles
located on the same site is $ 4\lambda\cos (\gamma n+\beta )+U $. The
transition
amplitude on the diagonal $ n_{1,2}=n $ is given by the amplitude of the
hopping via virtual states with $ n_1-n_2=\pm 1 $ and energy denominator $1/U$.
There are two such paths so that the effective amplitude
is $V_{\rm eff} = 2/U$. The same expression can be derived by the Schur formalism
(see Sec. IV). After dividing the Hamiltonian by $V_{\rm eff}$
we arrive to the eigenfunctions equation in the form of Harper (14) with $
\lambda $ replaced by $ \lambda _{\rm eff} =U\gg 1 $. Since $ \lambda _{\rm
eff}\gg 1 $  when $ U\gg 1 $, the pair states are always within the
localized phase of the Harper equation showing exponential localization. In
Fig. 10, we show a typical eigenstate of the Mathieu part of the spectrum
for $ U=50 $ and $ \gamma /2\pi =34/55 $. The fact that it is localized
confirms the pure-point character of the corresponding spectrum.

Above we showed that in the case of strong interaction, we have $ \lambda _{\rm eff}\gg
1 $. This explains the appearance of a pure-point component in the spectrum.
However, we think that this pure-point component will even appear for small
values of the interaction. Our argument is the following : without interaction,
the system obeys Aubry's duality while the presence of the interaction
introduces Aubry's duality breaking. Indeed, from (18) it is easy to see that
the interaction acts in the coordinate space and the symmetry with momentum
space disappears when $ U\neq 0 $. Formally, this argument is not sufficient to
prove the existence of pure-point spectrum at arbitrary small $ U $. However
the
ensemble of numerical datas we have here and in \cite{Shepe96,BaBeJaSh}
confirms this conjecture.

\vspace{-20mm}

\let\picnaturalsize=N
\def\picsize{4.0in}
\def\picfilename{fig10.eps}
\ifx\nopictures Y\else{\ifx\epsfloaded Y\else\input epsf \fi
\let\epsfloaded=Y
\centerline{\ifx\picnaturalsize N\epsfxsize \picsize\fi
\epsfbox{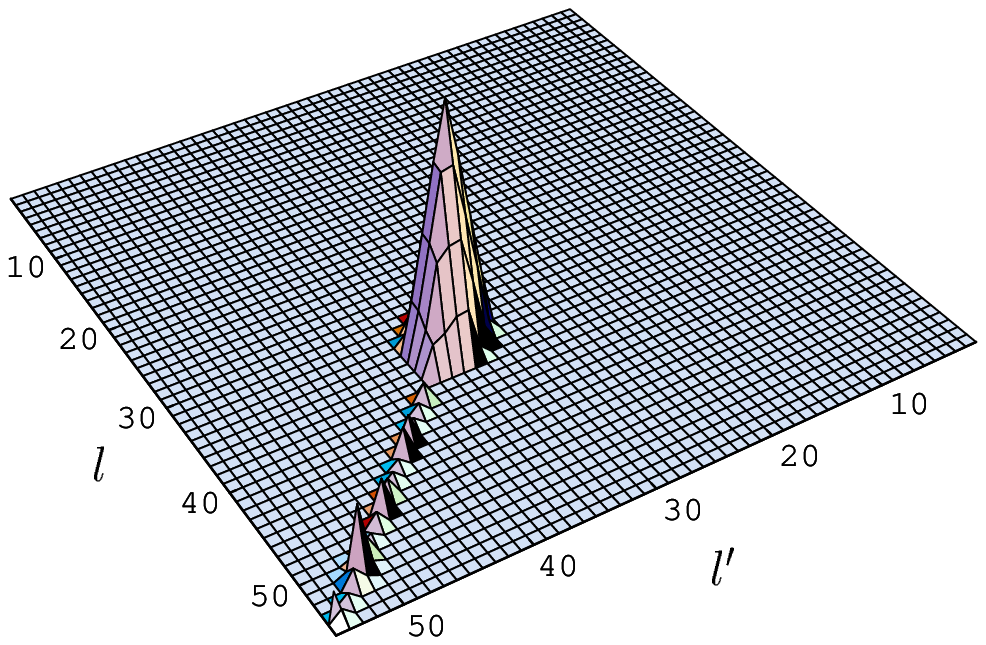}}}\fi

FIG. 10. Semilogplot of $ W=\vert\phi (l,l')\vert ^2 $ for a localized state
(E=50.25,$-30\leq\ln W\leq -1$).

\vspace{3mm}

When $ U $ is large, the unshifted part of the spectrum looks very much like
the spectrum at $ U=0 $. The main difference can be found by looking carefully
at the Landau levels (see Fig. 11). The
reminiscence of the existence of the interaction is seen through the appearance
of a splitting of Landau sublevels. This splitting only exists when Landau
quantum numbers are different $ \nu _1\neq \nu _2 $ and the two particles are
located in the same well.
Such a behaviour is illustrated by Fig. 11. The other sublevels are described
by the semiclassical formulae obtained in the case $ U=0 $ (25).
To derive this splitting using semiclassical analysis, we again apply the
Schur complement formula. Dealing with the unshifted butterfly leads us to
consider as the dominant term $ QH_{\rm TIP}Q $ such that (37) becomes :
\begin{equation}
H_{\rm TIP}^{{\rm eff}}(z)=Q H_{\rm TIP}Q +QH_{\rm TIP}P\frac{1}{z-PH_{\rm TIP}
P}P H_{\rm TIP}Q
\end{equation}

Applying the same scheme as before produces an additional shift from the
unperturbed energy given in first order in $\gamma$ by :
\begin{equation}
\vert\delta E (\gamma)\vert = 4 \frac{\gamma}{U+4}
\end{equation}
This shift is valid for the second Landau sublevel ($ \nu _1=0,\nu _2=1 $),
its accuracy is shown in Fig. 11 and the two splitted subbands are given by :
$ E(\gamma )=8-4.1666\gamma $ and $ E(\gamma )=8-4\gamma $ up to order 1 in $
\gamma $.

\vspace{-5mm}

\let\picnaturalsize=N
\def\picsize{4.0in}
\def\picfilename{fig11.eps}
\ifx\nopictures Y\else{\ifx\epsfloaded Y\else\input epsf \fi
\let\epsfloaded=Y
\centerline{\ifx\picnaturalsize N\epsfxsize \picsize\fi
\epsfbox{\picfilename}}}\fi

\vspace{-1cm}
FIG. 11. Semiclassical calculations (45) (full curve) and exact numerical spectrum
(points) for the splitting of the $\nu_1=0, \nu_2=1$ Landau sublevel in the
two-particle Harper model for $ U=20 $.

\section{Two interacting particles on a two-dimensional lattice}
Even though the studied model was derived from a model of two-dimensional
electrons, its effective dimension is 1 : as we already pointed out, (18) was
derived assuming that the particle propagate as plane-wave in one direction.
This assumption, though reasonable in the one-particle model, could be
violated by interaction induced quantum interferences in the two-particle case.
Therefore the question of the survival of interaction induced localization 
effect for two interacting
particles in two dimensions remains an open problem. In this section we would
like to discuss briefly this situation. For two interacting particles moving on
a
two-dimensional square lattice submitted to a uniform magnetic flux, the
eigenvalue
equation reads :
\begin{eqnarray}
{\rm e}^{i\gamma y_1}\psi_{x_1 +1, y_1, x_2, y_2} +
{\rm e}^{-i\gamma y_1}\psi_{x_1 -1, y_1, x_2, y_2}\nonumber\\  
+\psi_{x_1, y_1 +1, x_2, y_2}+ \psi_{x_1, y_1 -1, x_2, y_2}\nonumber\\  
+{\rm e}^{i\gamma y_2}\psi_{x_1, y_1, x_2 +1, 
y_2} +{\rm e}^{-i\gamma y_2}\psi_{x_1, y_1, x_2 -1, y_2}\nonumber\\
+\psi_{x_1, y_1, x_2, y_2 +1} + \psi_{x_1, y_1, x_2, y_2 -1}\nonumber\\   
+\tilde{U} \delta_{x_1, x_2} \delta_{y_1, y_2} \psi_{x_1, y_1, x_2, y_2}
= E \psi_{x_1, y_1, x_2, y_2}
\end{eqnarray}
where $ (x_{1,2},y_{1,2}) $ are integers denoting the positions on the square
lattice and $ \tilde{U} $ is the on-site interparticle interaction. For $
\tilde{U} =0 $,
the previous equation can be reduced to the one-dimensional Harper equation we
discussed above (18). With interaction, the same equation (18) can be obtained
in the ansatz of plane waves propagating in one direction with renormalized
interaction $ U $ \cite{Shepe96}. While this plane wave approximation is a
standard
approach for the one-particle Harper problem, it has to be handled with care in
the interacting case. Indeed this plane wave ansatz breaks the symmetry of the
original
problem (46). This symmetry can be seen in the limit of strong interaction $
U\gg 1 $. In this case, there should be two energy bands : one corresponding to
the pair states when particles are located on the same site with energy $
E\approx U $ and the other with $ E\approx 1 $ for the states in which the two
particles avoid each other. In the higher energy band, the eigenvalue equation
for the pair states up to the terms of order $ 1/U $ has the form :
\begin{eqnarray}
\displaystyle\frac{2}{U}\left({\rm e}^{2i\gamma y}\phi_{x +1, y} + 
{\rm e}^{-2i\gamma y}\phi_{x -1,
y} +\phi_{x, y +1} + \phi_{x, y -1}\right)\nonumber\\ +
U\phi_{x,y} = E\phi_{x, y}
\end{eqnarray}
Here the term $ 2/U $ represents the transition amplitude for pair states. Its
derivation is similar to the case of two interacting particles in the 
one-dimensional Harper model. Indeed if one keeps $ x_1=x_2 $ then the hopping
term is given by $ V_{\rm eff}=2/U $ because there are two paths with
virtual energy $ U $ ($ y_{1,2}
\rightarrow y_{1,2}+1 $) which contribute to the hopping term in the $ y $-
direction. Similarly the hopping in the $ x $-direction is $ V_{\rm eff}=2{\rm
e}^{\pm 2i\gamma }/U $. 

This representation shows that the symmetry between the two directions or the
Aubry duality is not broken by the interaction. The main reason is that the
symmetry of the interaction is invariant under rotations on the square lattice.
In the limit of large $ U $, this property can be seen through equation (47).
However the symmetry (Aubry's duality) should also be preserved for small
interaction. Due to that, we expect that similarly to the Harper model with $
\lambda =1 $, the interaction will not generate pure-point component in the
spectrum. However this conjecture has to be directly checked in further
analytical and numerical studies.

\section{Conclusions}
In this paper we have emphasized a localizing effect due to the combined
action of an on-site interaction and a quasiperidodic potential. Unlike in the
random potential case \cite{TIP}, extended unperturbed states are localized by
the interaction, and this localization occurs at arbitrarily small
attractive/repulsive interaction. We successfully identified the mechanism
responsible for this effect as a decoupling of a Mathieu-like model from the
original two-particle Harper model in the limit of large interaction. Our
conjecture is that a similar mechanism will also work for small $ U $ due to 
an interaction induced breaking of Aubry's duality. This breaking happens in
one-dimensional incommensurate models, however in two-dimensional magnetic
models, we expect that the interaction will not break the duality and that a
pure-point component in the spectrum will not arise. Further verifications of
these conjectures are required.

\vspace{3mm}

This work has been supported in part by the Fonds National Suisse de la
Recherche. Two of us (A.B. and Ph.J.) want to thank the Institut de Physique Neuch\^atel
(Switzerland) and the Laboratoire de Physique Quantique Toulouse (France) for 
hospitality.

\end{document}